\documentclass[twocolumn,secnumarabic,superscriptaddress,amssymb, nobibnotes, aps, prl]{revtex4-1}

\usepackage{comment}
\usepackage[dvipsnames]{xcolor}
\usepackage{empheq}
\usepackage{amsmath}
\usepackage{bbold}
\usepackage{amssymb}
\usepackage{ulem,xpatch}
\usepackage[paperwidth=210mm,paperheight=297mm,centering,hmargin=2cm,vmargin=2.5cm]{geometry}
\usepackage{tabularx}

\usepackage{siunitx}

\usepackage{appendix}

\newcommand{\hide}[1]{\relax}

\newcommand{\Om}{\ensuremath{\Omega_\mathrm{m}}}

\newcommand{\Gm}{\ensuremath{\Gamma_\mathrm{m}}}

\newcommand{\mhat}{}

\newcommand{\ha}{\ensuremath{\mhat a}}
\newcommand{\had}{\ensuremath{\mhat a^\dagger}}

\newcommand{\hb}{\ensuremath{\mhat b}}
\newcommand{\hbd}{\ensuremath{\mhat b^\dagger}}

\newcommand{\nocontentsline}[3]{}
\newcommand{\tocless}[2]{\bgroup\let\addcontentsline=\nocontentsline#1{#2}\egroup}

\begin{document}

\title{A Long-lived and Efficient Optomechanical Memory for Light}%
\author{Mads Bjerregaard Kristensen}
\author{Nenad Kralj}
\altaffiliation{current address: Institute for Gravitational Physics, Leibniz Universität Hannover, Callinstraße 36, 30167 Hannover,
Germany}
\author{Eric Langman}
\author{Albert Schliesser}
	\email[to whom correspondence should be addressed: ]{\\albert.schliesser@nbi.ku.dk}
	\affiliation{Niels Bohr Institute, University of Copenhagen, 2100 Copenhagen, Denmark}
	\affiliation{Center for Hybrid Quantum Networks (Hy-Q), Niels Bohr Institute, University of Copenhagen, 2100 Copenhagen, Denmark}

\begin{abstract}
We demonstrate a memory for light based on optomechanically induced transparency. 
We achieve a long storage time by leveraging the ultra-low dissipation of a soft-clamped mechanical membrane resonator, which oscillates at MHz frequencies.
At room temperature, we demonstrate a lifetime $T_1 \approx \SI{23}{\milli\second}$ and a retrieval efficiency $\eta \approx 40\%$ for classical coherent pulses.
We anticipate storage of quantum light to be possible at moderate cryogenic conditions ($T\approx 10\,\mathrm{K}$).
Such systems could find applications in emerging quantum networks, where they can serve as long-lived optical quantum memories by storing optical information in a phononic mode.
\end{abstract}

\maketitle

\section{Introduction}

Optical quantum memories \cite{Lvovsky:2009aa} are an essential prerequisite for many exciting prospects of quantum information science, such as quantum cryptography \cite{Gisin:2002aa} over long distances and quantum-coherent interconnection of quantum computers \cite{Kimble:2008aa}.
Pioneering theoretical and experimental work in the atomic physics community has studied light storage in long-lived internal states of atoms, notably via electromagnetically induced transparency (EIT) \cite{Fleischhauer:2000aa}.
In EIT, an optical control field dynamically activates the conversion of an optical signal to an atomic excitation, and vice versa, with the control field switched off during storage.
For the specific purpose of entanglement distribution, the Duan-Lukin-Cirac-Zoller~(DLCZ) protocol \cite{Duan:2001aa} 
presents an alternative. 
However, due to the probabilistic nature and internal generation of excitations, it cannot be used to read an arbitrary optical state into a memory for later retrieval. 

Optical quantum memories building on these (and related) protocols have been implemented in a range of atomic systems that include hot and cold atomic vapors, single atoms and ions, as well as solid-state crystals doped with rare-earth ions (see \cite{Lvovsky:2009aa} and references therein). 
One disadvantage of atomic and ionic memories for light is that they are tied to the natural transition frequencies of the employed species.
This is often imcompatible with desired operation wavelengths, e.g. within a telecom band where optical losses in fibers are low.
Whereas this shortcoming can in principle be addressed through non-linear optical frequency conversion \cite{Zaske2012}, this comes at a significant experimental overhead in terms of cost, space and power.

Optomechanical systems \cite{Aspelmeyer:2014aa} offer an alternative implementation of optical quantum memories, the properties of which can be freely engineered.
They provide a coherent, wavelength-versatile interface between light and a resonant mechanical mode.
As a stationary quantum system with a potentially long coherence time, the latter acts as the memory.
Optomechanical memories employing the DLCZ protocol have indeed been proposed \cite{Galland2014} and demonstrated recently \cite{Wallucks:2020aa, Fiaschi:2021aa}, achieving coherence times of \numrange{10}{100} \si{\micro\second} and lifetimes $T_1 \approx \SI{2}{\milli\second}$
at millikelvin temperatures.
\begin{figure}[t]
    \centering
    \includegraphics[width=\columnwidth]{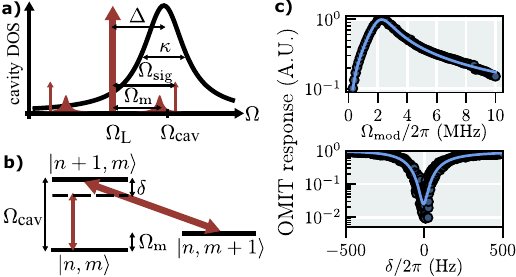}
    \caption{Steady-state OMIT. 
    a) An optomechanical cavity mode with frequency $\Omega_\mathrm{cav}$ and linewidth $\kappa$ is pumped near the red sideband $\Delta = \Omega_\mathrm{L}-\Omega_\mathrm{cav} \approx - \Omega_\mathrm{m}$. 
    An externally applied modulation at $\Omega_\mathrm{sig}$ provides a signal to store.
    b) The optical ($n$) and mechanical ($m$) excitations form a $\Lambda$-system which can facilitate storage when the two-photon resonance condition $\delta = \Omega_\mathrm{sig}-\Omega_\mathrm{m}\approx0$ is met.
    c) Experimental data (circles) for coarse, broadband (top) and fine, narrowband (bottom) sweeps of $\Omega_\mathrm{mod}$, with fits (solid blue) to OMIT theory. 
    The broad sweep outlines the cavity response, while the narrow sweep shows interference between the mechanical sideband and applied signal around $\delta = 0$.}
    \label{fig:0}
\end{figure}
Conveniently, optomechanical systems also exhibit a close analog to atomic EIT, known as  optomechanically induced transparency (OMIT) \cite{Weis:2010aa}, see fig.~\ref{fig:0}.
OMIT results from the beamsplitter-type process
\begin{equation}
  H_\mathrm{int} = - \hbar g_0  \alpha \left( \ha \hbd + \had \hb \right),
\label{eq:Hint}
\end{equation} which describes interconversion of optical (ladder operators \ha, $\had$) and mechanical (ladder operators \hb, $\hbd$) excitations.
This process dominates optomechanical interactions when the system is pumped by a  coherent field (amplitude $\alpha$) at the red motional sideband, and the optical cavity linewidth $\kappa$ is smaller than the mechanical resonance frequency $\Om$, the so-called resolved-sideband regime.
The vacuum optomechanical coupling rate $g_0$ is fixed by geometry and material properties.
It was recognized early on \cite{Schliesser2009, Weis:2010aa, Safavi2011} that  OMIT enables dynamic control of a storage and retrieval process by tuning the coherent amplitude $\alpha$, just as in EIT.
OMIT-based memories have  been demonstrated in silica microspheres \cite{Fiore:2011aa} and diamond microdisks \cite{Lake2021}, albeit with relatively short memory lifetimes (µs-timescales) and without quantifying storage efficiency.
Related experiments with propagating acoustic phonons suffer from even shorter lifetimes \cite{Stiller:2020aa}.
Millisecond-level storage times have yet only been demonstrated with microwave (instead of optical) pulses stored in mechanical systems held at millikelvin temperatures \cite{Palomaki:2013aa}.

In this work, we present an OMIT-type optical memory that operates at telecom-wavelengths. 
It is based on a long-lived mechanical mode of a soft-clamped membrane resonator \cite{Tsaturyan:2017aa} embedded in a Fabry-Pérot cavity \cite{Jayich2008}.
We characterize and model the performance as an optical memory by injecting coherent state signals and  measuring retrieval efficiency as a function of bandwidth, two-photon detuning and storage time. Ultimately, we demonstrate the longest lifetime and highest efficiency of any optomechanical memory to date.

\section{The experimental platform}

\begin{figure}[h]
    \centering
    \includegraphics[width = \columnwidth]{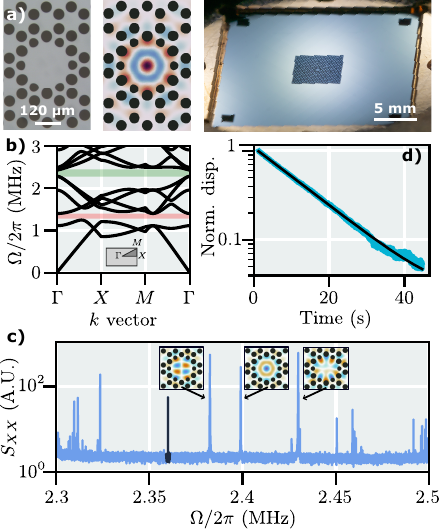}
    \caption{ Soft-clamped membrane resonator. 
a) The mechanical resonator. From left to right, a micrograph of the device used, a simulated displacement field of the mode we work with, and a photograph of an exemplary mechanical resonator.
b) In black, simulated band diagram, showing a fundamental (red) and higher-order (green) bandgap; inset: first Brillouin zone.
c) Mechanical displacement spectrum around the second bandgap, showing distinct localized modes, annotated with displacement profiles (the black peak at $\SI{2.36}{\mega\hertz}$ is an external phase modulation).
d) Amplitude-decay of the mode of interest, exhibiting $Q = \num{100\pm 3e6}$  at $\Omega_\textrm{m}/2\pi = \SI{2.4}{\mega\hertz}$.
}
    \label{fig:1}
\end{figure}

Fig.~\ref{fig:1} shows the mechanical memory, which is based on a $\SI{20}{\nano\meter}$ thin membrane of stoichiometric LPCVD $\textrm{Si}_3\textrm{N}_4$, with a large in-plane stress ($\Bar{\sigma} \approx \SI{1.15}{\giga\pascal}$ before release).
It is patterned with a honeycomb phononic crystal lattice, resulting in several bandgaps for out-of-plane vibrational modes \cite{Tsaturyan:2017aa}.
A central defect breaks the crystal symmetry and spatially localizes resonance modes.
Careful design of the defect geometry places the desired defect mode frequency in the second bandgap, allowing a higher frequency that well fulfills the resolved-sideband condition $\Om >\kappa$.

Thanks to the phenomena of dissipation dilution and soft clamping \cite{Tsaturyan:2017aa}, the defect modes of such membranes have very low dissipation rates $\Gm$ and correspondingly high Q-factors $Q=\Om/\Gm$.
In fig.~\ref{fig:1}d, we show a room-temperature ringdown measurement demonstrating $Q = \num{100\pm 3e6}$ at $\Omega_\mathrm{m}/2\pi = \SI{2.4}{\mega\hertz}$.
Working with a localized mode in the higher-order bandgap enables a high mechanical frequency without compromising the size of the working area.

\begin{figure}[b]
    \centering
    \includegraphics[width = \columnwidth]{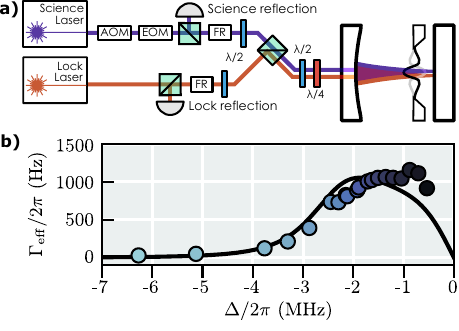}
    \caption{Membrane-in-the-middle cavity optomechanics. a) Simplified experimental schematic. All beamsplitters are polarizing. b) Mechanical memory bandwidth $\Gamma_\textrm{eff}$ as a function of science laser detuning $\Delta$ at fixed input power. The black line is a single free parameter ($g_0$) fit to a model of dynamical backaction.
    }
    \label{fig:2}
\end{figure}

We place the membrane inside a single-sided Fabry-P\'{e}rot cavity of length $L_\textrm{cav} \approx \SI{5.5}{\milli\meter}$ and high finesse $\mathcal{F} = 13600$, with a corresponding linewidth $\kappa/2\pi=\SI{2.1}{\mega\hertz}$.
By positioning the membrane between a node and an antinode of the standing optical wave, the ``membrane-in-the-middle''  linearly shifts the cavity resonance frequency upon displacement, leading to the standard dispersive optomechanical coupling 
\cite{Aspelmeyer:2014aa,Jayich2008}.

To use the membrane as a memory, we couple two lasers to the cavity, cf.\ Fig.~\ref{fig:2}. 
A strong 
``{science}'' laser, after being frequency-shifted with an acousto-optic modulator (AOM), provides the ``control'' field of amplitude $\alpha$.
It is red-detuned from a cavity mode, and drives the process of eq. (\ref{eq:Hint}) that facilitates the conversion between photons and phonons.
A second, weaker, ``{lock}'' laser is stabilized to the science laser at a frequency offset near the cavity free spectral range $\Delta\nu_\textrm{offset} \approx \nu_\textrm{FSR}$. 
The cavity, in turn, is stabilized to the lock laser with a Pound-Drever-Hall scheme. 
That is,  the two lasers address different longitudinal cavity modes, and additionally have orthogonal polarizations.
This setup allows us to stabilize the control field to detunings $-\Delta\approx \Om > \kappa$ outside the cavity bandwidth, as required for sideband-resolved optomechanics.
Furthermore, using the AOM, we can switch the control field off and on again at the same detuning, crucially only having the weaker lock laser on during the storage.
The experimental apparatus is described in more detail in the supplementary information.

We measure the effective memory bandwidth by recording mechanical noise spectra as we change the science laser detuning $\Delta$.
For the effective mechanical linewidth $\Gamma_\mathrm{eff}$ we observe fair agreement with a dynamical back-action (DBA) model that accounts also for the cavity susceptibility, cf. fig. \ref{fig:2}b (more details in supplementary). 
The largest optical broadening corresponds to a classical coorperativity $\mathcal{C} = 4 g_0^2 \alpha^2/\kappa \Gamma_\mathrm{m} > \num{4e4}$.
By simultaneously measuring the detuning, transmitted power and the optical broadening, we can extract $g_0/2\pi = \SI{1.0(1)}{\hertz}$ from the DBA fit.


\section{Light storage and retrieval}

\begin{figure}[ht!]
    \centering
    \includegraphics[width = \columnwidth]{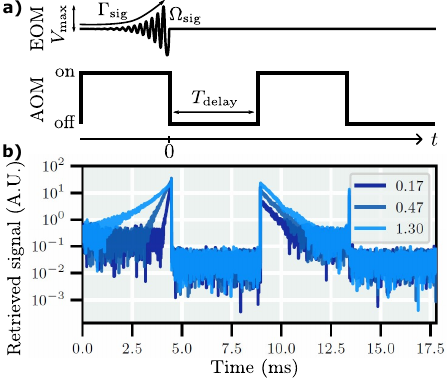}
    \caption{ Optomechanical light storage. 
    a) Pulse sequence diagram. We drive the EOM with exponentially rising sinusoids, while the science laser is on. 
    After waiting a time $T_\mathrm{delay}$, we switch the AOM on again and look at the retrieved mechanical signal.
    b) By lock-in detection around the mechanical sideband, we record the magnitude of the reflected science laser. Here we show data for fixed $\Gamma_\mathrm{eff}$ and increasing $\Gamma_\mathrm{sig}$ from light to dark blue, the legend showing $\Gamma_\mathrm{eff}/\Gamma_\mathrm{sig}$. }
    \label{fig:3}
\end{figure}

To characterize the optomechanical quantum memory, we perform several experiments with optical input pulses, see fig.~\ref{fig:3}.
Using the AOM, the control field is switched on for a first (write) and second (read) pulses of 
variable millisecond duration, separated by a storage time $T_\mathrm{delay}$, during which the control field is off.
During the write pulse, an electro-optic modulator (EOM) generates a ``{signal}'' field as a sideband of the control field.
The signal field falls on cavity resonance, and therefore to the center of the OMIT window. 
This is the field that will be stored as a mechanical excitation.
We choose exponentially rising sinusoid waveforms
\begin{equation}
    V_\mathrm{EOM}(t) = V_\mathrm{max} \sin(\Omega_\mathrm{sig} t) e^{(\Gamma_\mathrm{sig}t/2)}H(-t)
\end{equation}
with programmable amplitude $V_\textrm{max}$, bandwidth $\Gamma_\textrm{sig}$, frequency $\Omega_\textrm{sig}$, and the Heavyside step function $H(t)$.
This gives rise to a signal field 
\begin{equation}\label{eq:sin}
    s_\mathrm{sig}^\mathrm{in}(t) = i\beta s_0 \sin(\Omega_\mathrm{sig} t)\mathrm{e}^{\Gamma_\mathrm{sig}t/2}H(-t),
\end{equation} 
impinging on the cavity. Here $\beta = \pi V_\mathrm{max}/ V_\mathrm{\pi}\ll 1$ is the modulation depth, $V_\mathrm{\pi}$ denoting the modulator half-wave voltage.

When the read control light pulse probes the cavity, the EOM is off.
Yet the control field acquires a sideband due to the mechanical motion, regenerating a signal field  $s_\mathrm{sig}^\mathrm{out}(t)$ at the cavity output.
A detector measuring optical power emerging from the cavity then retrieves a transient beat $V_\mathrm{ret}(t)\propto |\alpha^* s_\mathrm{sig}^\mathrm{out}(t)|$ oscillating at the modulation frequency $\Omega_\textrm{sig}$.
On the time scale of the state-swap process, the cavity responds fast, $g_0 \alpha \ll \kappa$, meaning we can adiabatically eliminate the mode $a$ and describe $s_\mathrm{sig}^\mathrm{out}$ in terms of $b$ using a Heisenberg-Langevin equation and input-output relation:
\begin{align}
     \Dot{b}(t) =& -\left(\frac{\Gamma_\mathrm{eff}}{2} - i \delta \right) b(t) +
    \sqrt{\eta_\mathrm{c} \Gamma_\mathrm{opt}} s_\mathrm{sig}^\mathrm{in}(t), \label{eq:QLE}\\
    s_\mathrm{sig}^\mathrm{out}(t) =& s_\mathrm{sig}^\mathrm{in}(t)  - i \sqrt{\eta_\mathrm{c} \Gamma_\mathrm{opt}} b(t).\label{eq:inout}
\end{align}
Here, $\Gamma_\mathrm{eff}=\Gamma_\mathrm{m}+\Gamma_\mathrm{opt}$ with  $\Gamma_\mathrm{opt}$ the optomechanical broadening, which is zero during the storage period. The signal input $s_\mathrm{sig}^\mathrm{in}$ is non-zero only during the write pulse, and $\delta = \Omega_\mathrm{sig}-\Omega_\mathrm{m}$ the two-photon detuning.
We can therefore obtain the magnitude of the retrieved signal field $|s_\mathrm{sig}^\mathrm{out}(t)|$ by demodulating the photodetector signal $V_\mathrm{ret}(t)$ at the frequency $\Omega_\textrm{sig}$ with a lock-in amplifier.
Exemplary time traces of such storage experiments are shown in fig.~\ref{fig:3}b. 

\begin{figure}[t]
    \centering
    \includegraphics[width = \columnwidth]{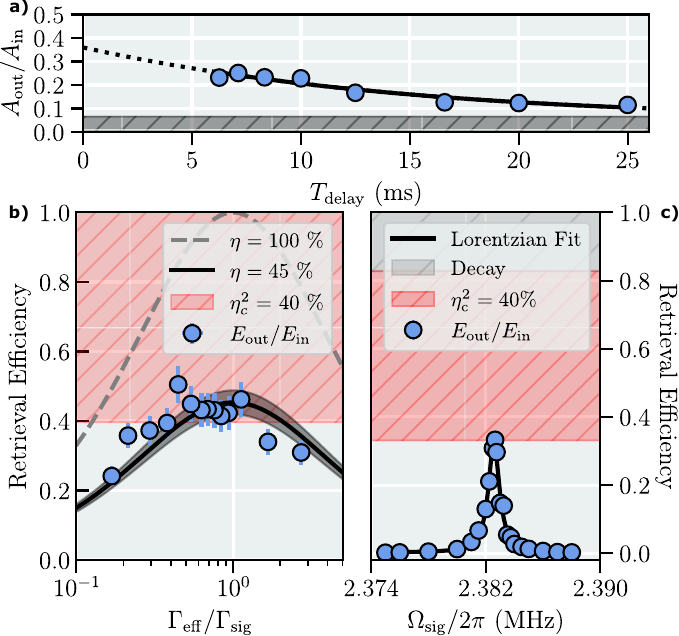}
    \caption{ Storage efficiency contributions. 
    a) Normalized retrieved amplitude subject to increasing readout delay. 
    The gray hatched area indicates the thermomechanical noise floor.
    b) Blue points: retrieved energy $E_\mathrm{out}$ normalized to signal energy $E_\mathrm{in}$ for varying $\Gamma_\mathrm{sig}$ and fixed $\Gamma_\mathrm{eff}$. In dashed gray, the theoretical prediction eq. \eqref{eq:etaBW} corresponding to unity efficiency.
    The black curve is a fit to the same model with one free multiplicative constant, with the shaded region indicating $\pm$ twice the error on the fit.
    The red hatched area indicates the limit to efficiency predicted by the cavity over-coupling, in fair agreement with the fit.
    c) Two-photon detuning. For fixed $\Gamma_\mathrm{sig} \approx \Gamma_\mathrm{eff}$ the signal frequency $\Omega_\mathrm{sig}$ is tuned across the mechanical resonance. The normalized, retrieved energy (blue points) is well-modelled by a Lorentzian, as predicted by eq.\eqref{eq:etadelta}.
    The gray hatched area indicates for the mechanical decay during the storage time.
    Accounting for the limitations of over-coupling, and the mechanical decay during the storage, we find good quantitative agreement between measurements and the expected bound.
    All errorbars indicate twice the standard error on the mean.}
    \label{fig:4}
\end{figure}

To quantify the retrieval efficiency $\eta$ of our memory, we systematically vary the signal field parameters in a range of experiments.
We define the memory efficiency $\eta\equiv E_\mathrm{out}/E_\mathrm{in}$ as the ratio of the time-integrated energy content of the retrieved field, $E_\mathrm{out}$, relative to that of the input field $E_\mathrm{in}$. 
Both the cavity over-coupling $\eta_\mathrm{c}$ and an internal efficiency $\eta_\mathrm{int}$ contribute to the memory efficiency as in $\eta=\eta_\mathrm{c}^2\cdot \eta_\mathrm{int}$, where the cavity overcoupling is squared accounting for the fact that light has to enter and leave the cavity.
Here, the cavity overcoupling $\eta_\mathrm{c}=63\%$ is limited by excess intracavity loss due to the membrane.
Upon detection, we incur additional losses due to light lost along the optical path ($\eta_\mathrm{loss}=60\%$) and the quantum efficiency of our photodetector ($\eta_\mathrm{QE}=83\%$), measured with a power meter and taken from the specifications respectively, totalling $\eta_\mathrm{det}=\eta_\mathrm{loss}\cdot \eta_\mathrm{QE} = 50\%$.
In the following we study the internal efficiency $\eta_\mathrm{int}$, cf. fig.~\ref{fig:4}.

We use time traces like those in fig. \ref{fig:3}b to measure the retrieved energy $E_\mathrm{out}\propto \int_{T_\mathrm{delay}}^\infty\left|V_\mathrm{ret} \right|^2 dt$. We estimate the signal energy  $E_\mathrm{in}\propto \int_{-\infty}^0\left|V_\mathrm{in} \right|^2 dt$, when the signal does not probe the mechanics but remains near cavity resonance, $\kappa \gg |\delta| \gg \Gamma_\mathrm{eff}$.

We first study the effect of mechanical decay, cf. fig. \ref{fig:4}a, by varying $T_\mathrm{delay}$. 
Ideally, this is only limited by the bare mechanical decay time, unmodified by dynamical backaction. 
In practice, the lowest power of the lock laser we can operate at is around $\approx \SI{2}{\micro\watt}$. Since we also slightly red-detune it to $\Delta_\mathrm{lock} \approx - 0.05 \kappa$ for stability reasons, it gives rise to a small amount of DBA-induced damping.
For each $T_\mathrm{delay}$ we compare our retrieved amplitude $A_\mathrm{out}$ to the drive signal amplitude $A_\mathrm{in}$ from a corresponding reference measurement.
In fig.~\ref{fig:4}a we plot these measurements and observe an exponential amplitude decay at a rate $1/2 T_1$ with  $T_1 \approx \SI{23}{\milli\second}$. We attribute this increase over the intrinsic value $\Gamma_\mathrm{m}/2$ to the presence of the lock laser, even \textit{in the dark}, when the science laser is off. We corroborate this by observing an decrease in $T_1$ with increasing lock laser power.
Consequently, in eq.~(\ref{eq:QLE}), we have $\Gamma_\mathrm{eff}\mapsto 1/T_1+\Gamma_\mathrm{opt}$.

Next we study the effect of bandwidth matching, by fixing $\Gamma_\mathrm{eff}$, setting $T_\mathrm{delay}=\delta=0$, and varying $\Gamma_\mathrm{sig}$, cf. fig. \ref{fig:4}b.
From equations (\ref{eq:sin},\ref{eq:QLE},\ref{eq:inout}) we calculate in this case 
\begin{equation}\label{eq:etaBW}
    \eta_\mathrm{int} = \frac{4 \Gamma_\mathrm{sig}\Gamma_\mathrm{eff}}{(\Gamma_\mathrm{sig}+\Gamma_\mathrm{eff})^2}.
\end{equation}
Fitting the data with the model of eq.~\eqref{eq:etaBW}, we obtain a multiplicative constant of $\eta=0{.}45 \cdot \eta_\mathrm{int}$, reasonably compatible with the expected $\eta_\mathrm{c}^2 = 40\%$.

Furthermore, we study the effect of two-photon detuning $\delta$, cf. fig. \ref{fig:4}c.
For a fixed delay $T_\mathrm{delay} \approx \SI{4.4}{\milli\second}$, and matched bandwidth $\Gamma_\mathrm{sig} \approx \Gamma_\mathrm{eff}$, we vary $\Omega_\mathrm{sig}$ and accordingly $\delta$. From eq. \eqref{eq:QLE} we expect the mechanics to respond with a Lorentzian susceptibility, leading to
\begin{equation}\label{eq:etadelta}
    \eta=\eta_\mathrm{c}^2 \cdot \frac{\Gamma_\mathrm{eff}^2}{\delta^2 + \Gamma_\mathrm{eff}^2} \cdot e^{-\frac{T_\mathrm{delay}}{T_1}}.
\end{equation}
We observe good agreement between our data and model, both at the qualitative and quantitative level.
Accounting for mechanical decay during storage, and the limit imposed by the overcoupling, we achieve the expected efficiency predicted by eq. \eqref{eq:etadelta}.

\section{Discussion}

Taken together, our measurements demonstrate a retrieval efficiency of $\eta \approx 40 \%$, with well-understood scalings. 
This is limited by intracavity optical losses.
We also achieve an already long $T_1 \approx \SI{23}{\milli\second}$ lifetime, limited by the smallest lock laser power we can operate with. 
Upgrading to a more sensitive detector should push this number closer to $\Gamma_\mathrm{m}^{-1}\approx \SI{6.3}{\second}$.
Taken together, this makes the present work the first simultaneous demonstration of a telecom wavelength, efficient, and long-lived OMIT-based memory for light.
In fact, our room-temperature experiment already shows the slowest decay time $T_1$ and highest efficiency $\eta$ of any optomechanical memory to date.

In addition, the prospects for quantum-coherent operation are promising. 
For a future cryogenic experiment we extrapolate a heating-limited coherence time of $1/\Gamma_\mathrm{m} \Bar{n}_\mathrm{th}\approx 0.2\, \mathrm{ms}$, conservatively assuming thermalization to $\SI{10}{\kelvin}$ and a three-fold improvement in $Q$-factor.
Furthermore, potential effects of pure dephasing should be investigated, even though our membrane resonators have previously not shown dephasing beyond that caused by heating by the thermal bath \cite{Rossi2018, Rossi2019}.
With modest improvements in coupling rate $g_0$ and optical losses $\kappa$, we envision storage and retrieval with larger bandwidth $\Gamma_\mathrm{eff}$, enabling shorter (sub-millisecond) optical pulses.

In terms of storage of quantum states of light, single-photon storage is of great interest due to the quantum network applications \cite{Kimble:2008aa}.
Based on our findings, the main limitation to efficiency in such an experiment would be the bandwidth ratio, where we find a $\eta \propto \Gamma_\mathrm{eff}/\Gamma_\mathrm{sig}$ scaling for $\Gamma_\mathrm{eff}\ll \Gamma_\mathrm{sig}$.
In this vein, a range of approaches for generating narrowband single photons is being pursued, with spontaneous parametric downconversion \cite{Rambach2016}, non-linear optical bandwidth compression \cite{Sosnicki2023}, and charged-exciton Raman transitions in quantum dot sources \cite{Beguin2018} as prominent examples.
Regarding the latter, recent progress in controlling the charge noise \cite{Uppu2020} and nuclear spin bath in such devices bodes well for transform-limited, temporally longer single photons \cite{Zaporski2023}, enabled by spin coherence times improved from nano- to micro-seconds.
Such sources would facilitate storage and retrieval of single photons in our OMIT memory.

\section*{Acknowledgements}
The authors acknowledge helpful discussions with Anders S. Sørensen regarding the theoretical modelling. 
This work was supported by the European Research Council project PHOQS  (grant no.~101002179), the Novo Nordisk Foundation (grant no.~NNF20OC0061866), the Danish National Research Foundation (Center of Excellence “Hy-Q”), as well as the Independent Research Fund Denmark (grant no. 1026-00345B).

%

\newpage

\onecolumngrid
\newpage
\appendix

\section{Supplementary Information}

\subsection{Mechanical Resonator}
\begin{figure}[h!]
    \centering
    \includegraphics[width = 0.75\columnwidth]{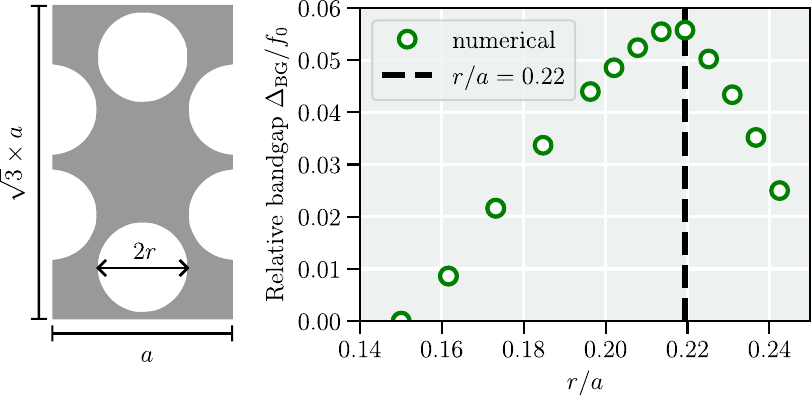}
    \caption{Unit cell design. Left: Parametrization of the unit cell. Right: Simulated relative second-order bandgap for varying hole radius.}
    \label{fig:SI:bg}
\end{figure}

Here we describe the design of the mechanical resonator.
First we optimize the phononic crystal to give the widest second-order bandgap. 
For a fixed lattice constant $a$ (see fig.~\ref{fig:SI:bg}~left), we sweep the hole radius $r$ and we find an optimal value of $r/a \approx 0.22$, cf.~fig.~\ref{fig:SI:bg}~right.

\begin{figure}[h!]
    \centering
    \includegraphics[width = 0.72\columnwidth]{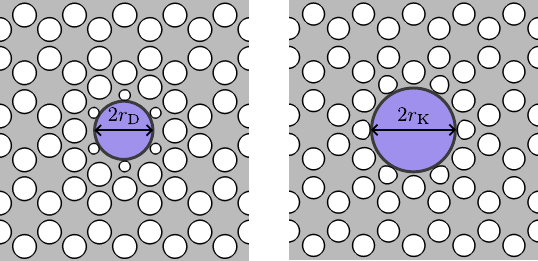}
    \caption{Defect design. Left we show the traditional ``Generation 2 Dahlia'' design, which was used in eg. \cite{Rossi2018}, with the defect size indicated by the largest inscribed circle of radius $r_\mathrm{D}$ shown in blue. Right we show the new design used in this work, exhibiting a significantly larger defect $r_\mathrm{K}\approx1.45 \times r_\mathrm{D}$, for the same lattice constant $a$. }
    \label{fig:SI:defect}
\end{figure}

With the phononic crystal design in place we turn to the defect. 
Our design goal is to simultaneously achieve a large defect and a high mechanical frequency $\Omega_\mathrm{m}$.
This is untenable by geometric rescaling of the device since $\Om \propto 1/a$ while $r_\mathrm{defect}\propto a$. 
Thankfully we can both utilize the fact the tensile stress in our perforated devices is anisotropic and that we have engineered a higher order band gap.
Tweaking the perforations near the defect lets us tailor the stress field, fine-tuning the defect mode frequencies, and the higher order bandgap gives us a factor $\approx 2$ over the first bandgap, for the same defect size.
In fig.~\ref{fig:SI:defect} we show the result of our optimization (right), with our so-far usual design for comparison (left).

\begin{figure}[h!]
    \centering
    \includegraphics[width=0.75\columnwidth]{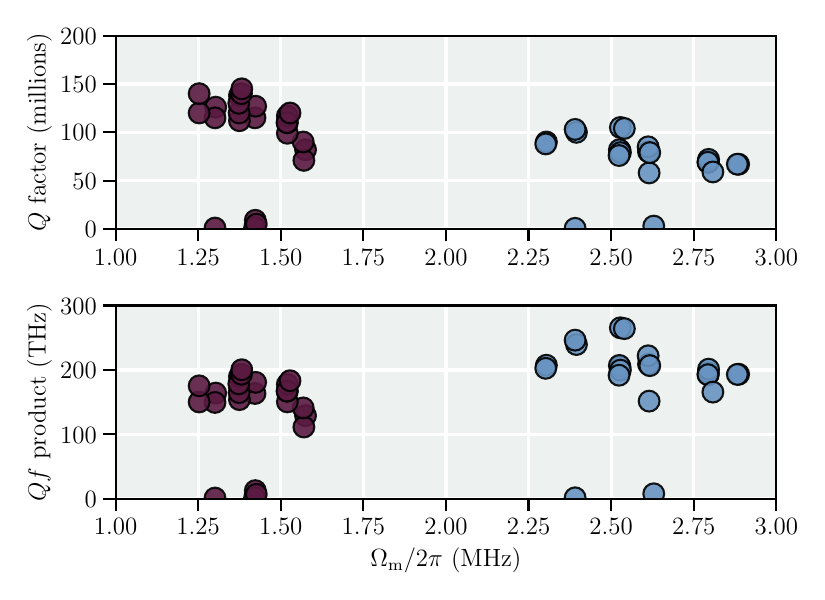}
    \caption{Top: $Q$ factors for the first (burgundy) and second bandgap (blue) modes of the ``K" design.
    Bottom: corresponding $Qf$-products, highlighting the advantage of the second bandgap mode. }
    \label{fig:SI:Q}
\end{figure}

The work presented in the main text was conducted using the new ``Generation K Sunflower" design.
We confirmed that the new design preserves the high $Q$-factors associated with soft-clamped membrane resonators by fabricating devices with a range of sizes and conducting ringdown measurements, with a sample measurement shown in the main text.
In figure \ref{fig:SI:Q} we catalogue measured $Q$ factors and $Qf$ products for 24 different membranes.

While both modes have high $Q$, the higher $Qf$ products shown for the second bandgap mode means we choose this mode for our experiments where simultaneous high frequency and $Q$ are important.

\newpage

\newpage
\subsection{Experimental Setup}
\begin{figure}[h!]
    \centering
    \includegraphics[width = \columnwidth]{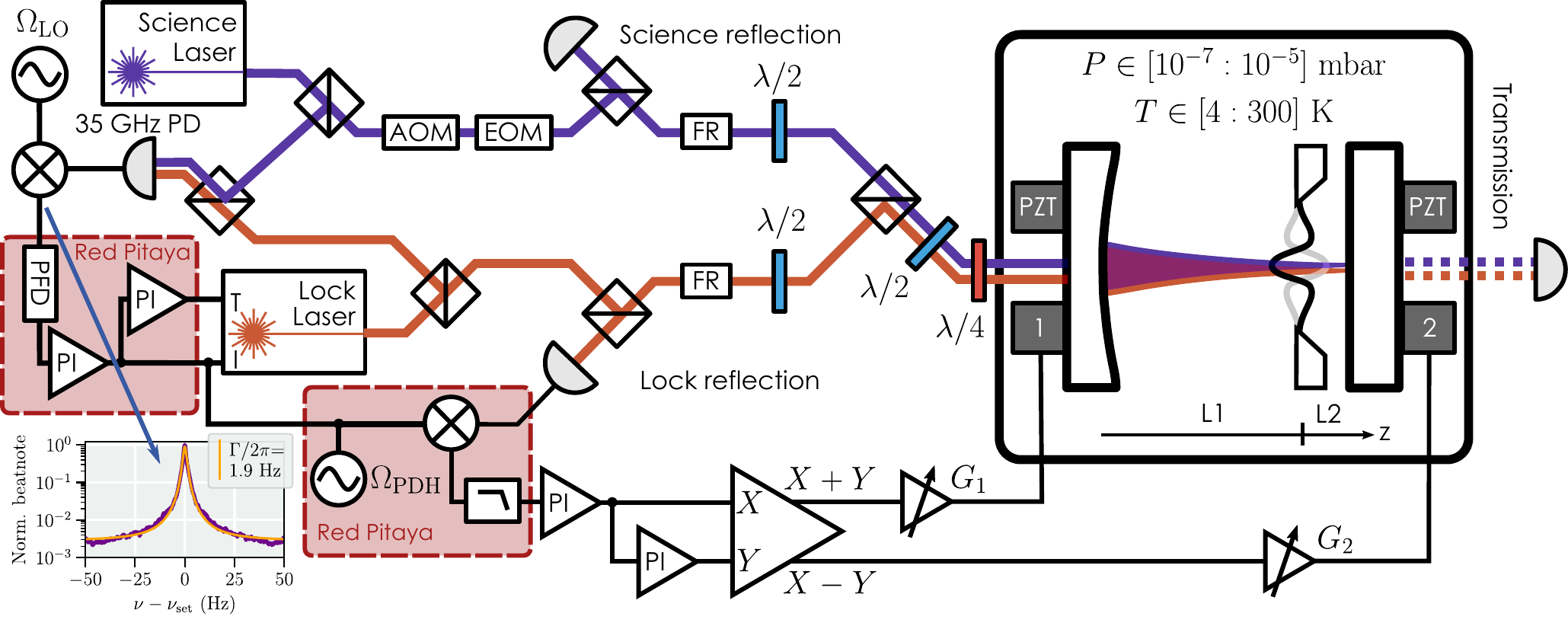}
    \caption{Detailed optical and electronic schematic of the experiment, see text for elaboration. AOM: acousto-optic modulator, EOM: electro-optic modulator, FR: Faraday rotator, $\lambda/2$: half-wave plate, $\lambda/4$: quarter-wave plate, PI: feedback controller with proportional and integral gain.}
    \label{fig:SI:setup}
\end{figure}

The optical and electronic elements of the experimental setup is shown in fig.~\ref{fig:SI:setup}.
As illustrated in fig.~\ref{fig:SI:setup}, we switch the science laser on and off using an acousto-optic modulator and inject the two lasers with orthogonal polarizations and directly detect the reflection of each laser independently, as well as their joint transmission. 
The lock laser reflection is processed for the PDH lock, whereas a phase-sensitive measurement of the science laser reflection is done with a lock-in amplifier.

A small fraction is tapped off each laser, such that we can measure their beat note on a fast  $\approx \SI{35}{\giga\hertz}$ bandwidth photodiode, which faithfully resolves the $\approx \SI{26}{\giga\hertz}$ beat note.
We downmix the measured beatnote to baseband and use a digital phase-frequency detector to derive an error signal that is PI processed and fed to the lock laser \cite{Prevedelli1995}.
The lock laser injection current (input $I$) has a high bandwidth (MHz) whereas the temperature (input $T$) counteracts long term drifts.
In this way we stabilize the beatnote of the two lasers to a full-width-half-max $\Gamma_\mathrm{beat}/2\pi \approx \SI{2}{\hertz}$.

With the two-lasers offset-locked we can PDH lock the cavity to the lock laser, and by extension effectively to the science laser.
This is implemented using a commercial analog lock-box, homebuilt sum- and difference-amplifier, and a commercial multichannel, high-voltage piezo amplifier with independently variable gains $G_1$ and $G_2$.
The two gains $G_1$ and $G_2$ are chosen such that they fulfill $G_1 L_1 = G_2 L_2$, with $L_1\approx \SI{4.7}{\milli\meter}$ and $L_2 \approx \SI{0.8}{\milli\meter}$ denoting the sub-cavity lengths.
In this manner the $2kz$ position is to first order insensitive to the common mode input $X$, which let's us stabilize the cavity length.
Experimentally we observed improved stability by augmenting the gain-balancing with a secondary slow (through the second integrator) feedback using the differential input.

\newpage
\subsection{Measuring $\Delta$, $\kappa$, and $g_0$}
\begin{figure}[h!]
    \centering
    \includegraphics[width = 0.61\columnwidth]{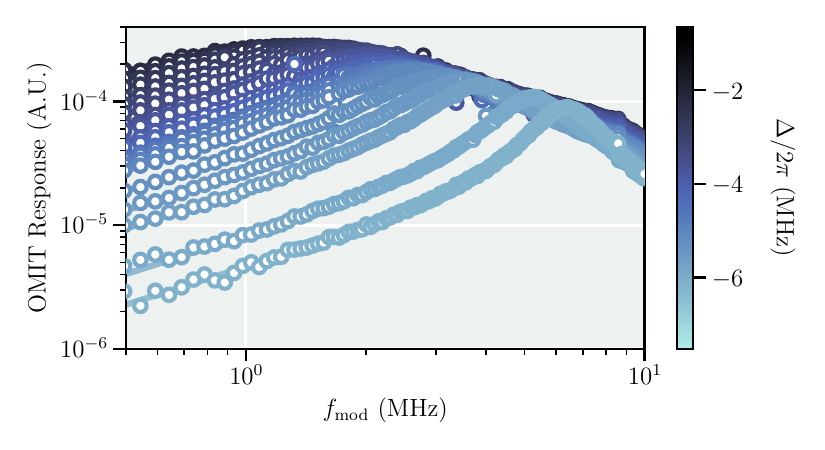}
    \caption{OMIT cavity spectroscopy. A fast and coarse sweep of a phase-modulation traces out the cavity response, enabling robust fitting from which we extract $\Delta$ and $\kappa$.}
    \label{fig:SI:OMIT}
\end{figure}

\begin{figure}[h!]
    \centering
    \includegraphics[width = 0.5\columnwidth]{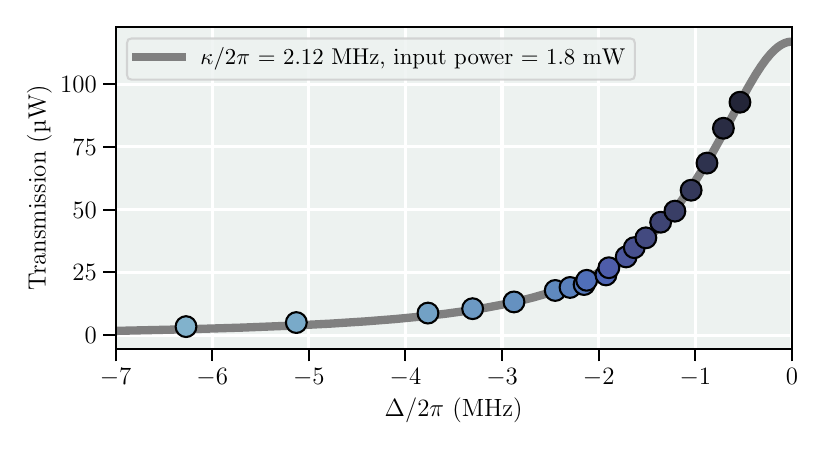}
    \caption{Transmitted power as a function of science laser detuning.
    The gray line is a one-parameter fit with a fixed linewidth from which we extract the input optical power.}
    \label{fig:SI:transmission}
\end{figure}

Here we describe our process to calibrate the parameters of our optomechanical cavity, namely detuning $\Delta$, linewidth $\kappa$, and vacuum coupling rate $g_0$.
For the detuning and linewidth we rely on OMIT sweeps, where the sweep step size is deliberately chosen $\delta_\mathrm{step} \gg \Gamma_\mathrm{eff}$.
In this way we can get a clean measurement of the bare cavity response, which we robustly can fit with an OMIT model with the coupling rate $g=0$.
This procedure is illustrated in fig.~\ref{fig:SI:OMIT} where hollow circles are measured data and solid lines are fits. 
The colours encode the extracted detuning as detailed on the color. 
This color-encoding is shared throughout the manuscript.
Each fit also give us a value for $\kappa$.
We take the average of these (the value quoted in the main text), and utilise it to extract the input power to our cavity. 
For each of the measurements in fig.~\ref{fig:SI:OMIT}, we also record the transmitted DC power level, which is shown in fig.~\ref{fig:SI:transmission}.
Using the OMIT-extracted cavity linewidth and knowledge of our mirror transmission coefficents ($\num{280e-6}$ and $\num{10e-6}$ respectively), we perform a single free parameter fit to the transmitted powers at a given detuning, from which we extract the input power.
With simultaneous knowledge of $P_\mathrm{in}$, $\eta_\mathrm{c}$, $\Delta$, $\Omega_\mathrm{m}$, and $\kappa$, we can fit our measured effective mechanical bandwidths $\Gamma_\mathrm{eff}$ to the following dynamical backaction model, as we vary the detuning \cite{Aspelmeyer:2014aa}:

\begin{subequations}\label{eq:gammaOptFit}
\begin{align}
     \Gamma_\mathrm{eff}(\Delta) &= \Gamma_\mathrm{m} + \Bar{n}_\mathrm{cav}(\Delta) g_0^2 \kappa \left(\frac{1}{(\kappa/2)^2 + (\Delta+\Omega_\mathrm{m})^2} - \frac{1}{(\kappa/2)^2 + (\Delta-\Omega_\mathrm{m})^2}\right), \\
     \Bar{n}_\mathrm{cav}(\Delta) &= \frac{P_\mathrm{in}}{\hbar \Omega_\mathrm{L}} \frac{ \eta_\mathrm{c} \kappa}{(\kappa/2)^2 + \Delta^2}.
\end{align}
\end{subequations}

With our other characterizations, only $\Gamma_\mathrm{m}$ and $g_0$ are left as free fit parameters. 
Since $\Gm$ is small we can further restrict the fit by setting $\Gm = 0$.
This is the model we use in fig.~\ref{fig:2} to extract a value for $g_0$.

\end{document}